\documentclass[a4paper,12pt]{article} 

\usepackage{amsmath}
\usepackage{color}
\usepackage{amssymb}
\usepackage{bm}
\usepackage{graphicx}
\begin{document}

\thispagestyle{empty}
\vspace*{-15mm}
{\bf OUJ-FTC-3}\\
{\bf OCHA-PP-356}\\
\renewcommand{\thefootnote}{\fnsymbol{footnote}}
\vspace{8mm}

\begin{center}
{\Large\bf
 Surround Inhibition Mechanism by Deep Learning\footnote{On the basis of this paper, Mamoru Sugamoto gave a talk at the 33rd Annual Conference of the Japanese Society for Artificial Intelligence (JSAI 2019), held at Niigata, Japan on June 5, 2019}}

\vspace{7mm}

\baselineskip 18pt
{\bf Naoaki Fujimoto${}^{1}$, Muneharu Onoue${}^{2}$, Akio Sugamoto${}^{3,}$${}^{4}$, Mamoru Sugamoto${}^{5,}$\footnote{mamoru.sugamoto@gmail.com},\\ and Tsukasa Yumibayashi${}^{6}$}

\vspace{2mm}

{\it ${}^{1}$Department of Information Design, Faculty of Art and Design, Tama Art University, Hachioji, 192-0394 Japan \\
${}^{2}$Geelive, Inc., 3-16-30 Ebisu-nishi, Naniwa-ku, Osaka, 556-0003, Japan\\
${}^{3}$Tokyo Bunkyo Study Center, The Open University of Japan (OUJ), \\
Tokyo 112-0012, Japan \\
${}^{4}$Ochanomizu University, 2-1-1 Ohtsuka, Bunkyo-ku, Tokyo 112-8610, Japan\\
${}^{5}$Research and Development Division, Apprhythm Co., Honmachi, Chuo-ku, Osaka, 541-0053 Japan\\
${}^{6}$Department of Social Information Studies, Otsuma Women's University, 12 Sanban-cho, Chiyoda-ku, Tokyo 102-8357, Japan}


\end{center}

\vspace{10mm}
\begin{center}
\begin{minipage}{14cm}
\baselineskip 16pt
\noindent
\begin{abstract}
In the sensation of tones, visions and other stimuli, the ``surround inhibition mechanism'' (or ``lateral inhibition mechanism") is crucial.  The mechanism enhances the signals of the strongest tone, color and other stimuli, by reducing and inhibiting the surrounding signals, since the latter signals are less important.  This surround inhibition mechanism is well studied in the physiology of sensor systems.  

The neural network with two hidden layers in addition to input and output layers is constructed; having 60 neurons (units) in each of the four layers.  The label (correct answer) is prepared from an input signal by applying seven times operations of the ``Hartline mechanism'', that is, by sending inhibitory signals from the neighboring neurons and amplifying all the signals afterwards. The implication obtained by the deep learning of this neural network is compared with the standard physiological understanding of the surround inhibition mechanism. 
\end{abstract}

\end{minipage}
\end{center}

\newpage
\section{Introduction}
The inhibitory influence between  different sensory receptors is found in the eye of Limulus (the horseshoe crab).  Limulus has a pair of compound eyes.  Each compound eye consists of 1000 ommatidia each of which is connected to a single nerve. Therefore, Limulus is the best creature to study the interactions between different light censors.

Hartline and his collaborators \cite{surround inhibition}, \cite{surround inhibition books} found by experiment that different ommatidia (light sensors) are interacted, giving mutual inhibition, that is, the excitation signal $ e_A$ of a sensor $A$ is inhibited from the neighboring sensor $B$, following the following linear equation:
\begin{eqnarray}
r_A=e_A-K_{AB} \times (r_B-r_B^0)~\theta(r_B-r_B^0), \\
r_B=e_B-K_{BA} \times (r_A-r_A^0)~\theta(r_A-r_A^0).  
\end{eqnarray}
Here, $e_{A, B}$ is the excitation at $A$ and $B$, induced when only $A$ or $B$ is stimulated. The $r_{A, B}$ is the response of the sensory nerve, when they are stimulated at the same time.  The $r_{A, B}^0$ gives thresholds, by a step function $\theta(r-r^0)~ (=1$ if $r-r^0>0$, but else $0$).  The strength of signal from a sensory nerve, is identified to the number of discharge of nerve impulses per unit time.

The inhibition (kinetic) coefficient $K_{AB}$ gives the rate of inhibition to the sensor $A$ from $B$.  If $A$ and $B$ are the equal type, the reciprocal law $K_{AB}=K_{BA}$ holds.  Its value depends on the pair of sensors, but is roughly $0.1-0.3$ by experiment.  In the later analysis, we use 0.25 for this inhibition coefficient.

Suppose that the sensors, labeled by $i$, are lined laterally, while the layer is labeled by $\ell$ to which the sensors are concentrated.  Then, Eq. (1) becomes
\begin{eqnarray}
r_i(\ell+1) =\sum_j r_i(\ell) -K_{ij} \dot (r-r^0)_j (\ell)~\theta(r_j-r_j^0)(\ell),
\end{eqnarray}
where we consider the response $r_j(\ell)$ be the input signal to a receptor $j$ located at layer $\ell$, and $r_i(\ell+1)$ is the output signal transferred to the receptor $i$ located at the next layer $\ell+1$. The 0-th layer can be the input layer, receiving the outside stimulus, $r_i(0)=e_i$.   

The same inhibition mechanism works for hearing.  In this case, signals are transferred from the cochlea where the voice sensor are placed, to cochlear nuclei, super olivary complex, inferior colliculus, lateral geniculate body (LGB), before arriving at the auditory cortex.  This means there exist at least four layers between the input and output, in which neurons are concentrated.

If we use the  ``rectified linear unit (ReLU)'' function defined by $y=f(x)=\mathrm{ReLU}(x)=x \times \theta(x-x^0)$, then we have
\begin{eqnarray}
x_i(\ell+1) =x_i(\ell) -\sum_j K_{ij} \times f\left((x-x^0)_j(\ell)\right).
\end{eqnarray}
This is identical to the neural network used by the deep learning (DL).  In DL the adjustable parameters are weights $W_{ij}$, which is identical to the (kinetic) coefficients $K_{ij}$ of Hartline {\it et al.}, \cite{surround inhibition}.  For the human hearing, the neural network having several layers are used.  

Now, we are ready to investigate by deep learning (DL) \cite{DL} how the sharpness of the contrast in brightness, color, or sound tone can be achieved. 

\section{Implication by Deep Learning (DL)}
We implement a neural network having two hidden layers, and the number of neurons (units) in four layers of input, hidden1, hidden2 and output is all 60.  The 1000 training data and the 100  test data are prepared by random separation from 2000 data.  Each datum and its label (a correct answer to be attained from a datum by DL) are prepared by the Harline's observation.  Given a one dimensional data of a certain signal (brightness or color of light, frequency of sound, or else), as a sum of sin functions:
\begin{eqnarray}
X(x)= \left(\sum_{n=1}^5 a_n \sin nx \right)^2,
\end{eqnarray}
where coefficients $0 \le \{a_1, \cdots, a_5\} \le 1$ are randomly generated, and $x$ is the discretized position of neurons, having 60 points $\{x_i=\frac{2\pi i}{60}  \vert i=0, \cdots, 59\}$. The ``Hartline operation'' $X'=\hat{H}X$ is defined by
\begin{eqnarray}
X'(i)=\lambda \cdot \mathrm{ReLU} \{X_i-\kappa(X_{i+1}+X_{i-1})\}.
\end{eqnarray}
The label $Y$ for a datum $X$ is given by $Y=(\hat{H})^7 X$ with $\kappa(=\mathrm{Hartline's}~K)=0.25$, and $\lambda=2$.

The running of DL for 100,000 epochs, using sigmoid as the activation function and the full Stochastic Gradient Descent method, yields the following epoch dependence of the loss function defined as the squared error, $L(X_{\mathrm{out}}, Y)=\frac{1}{2N} \sum_{n=0}^{N-1} \sum_{i=0}^{59} (X_{\mathrm{out}}-Y)^2$, where $N=100$ is the total number of test data.

See (Figure \ref{loss100,000}) which shows the decrease of loss function by epochs, as well as the intermediate epochs of DL training, corresponding to the Hartline mechanism.

{\begin{figure}[h]
\centering
\includegraphics[width=80mm]{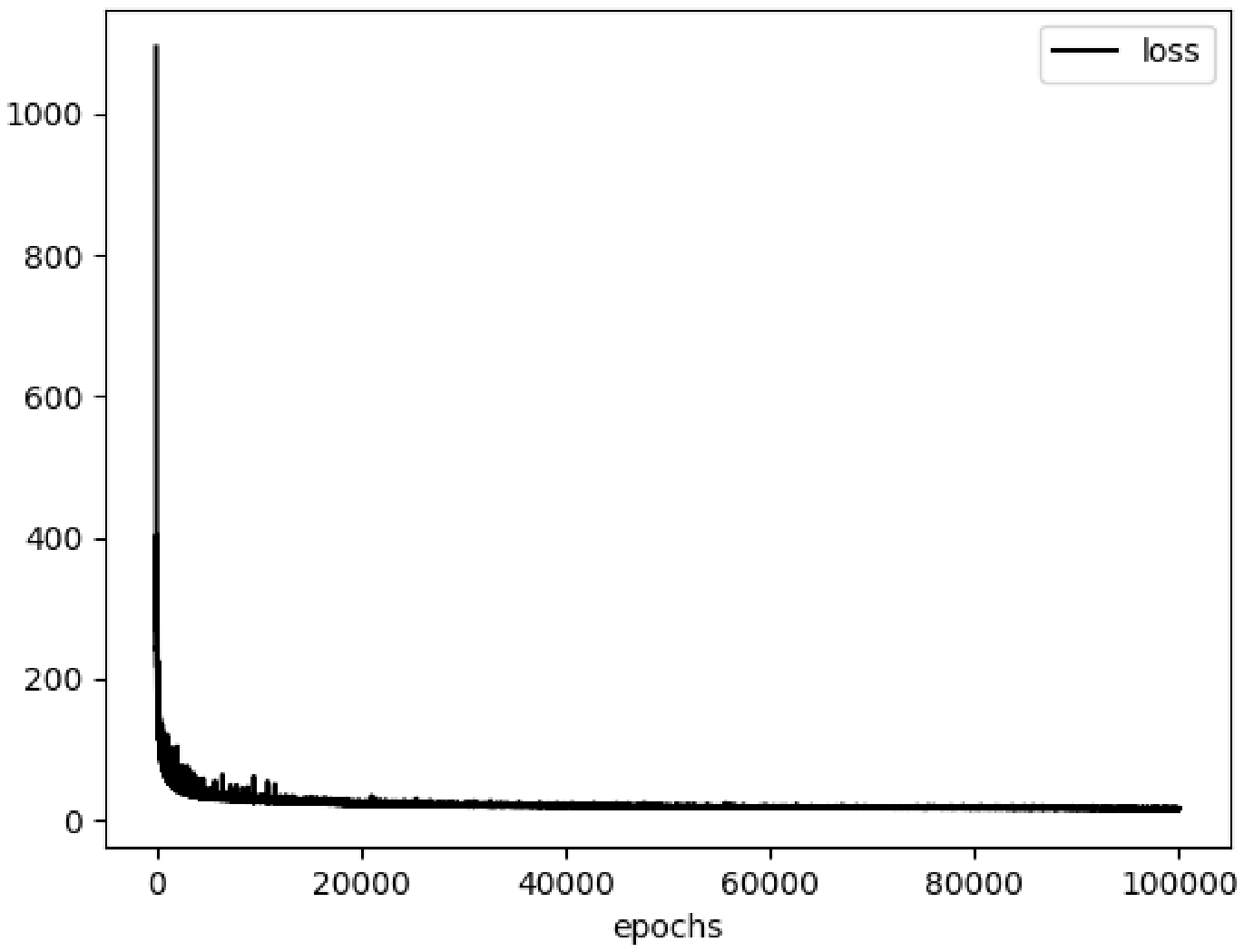}
\includegraphics[width=80mm]{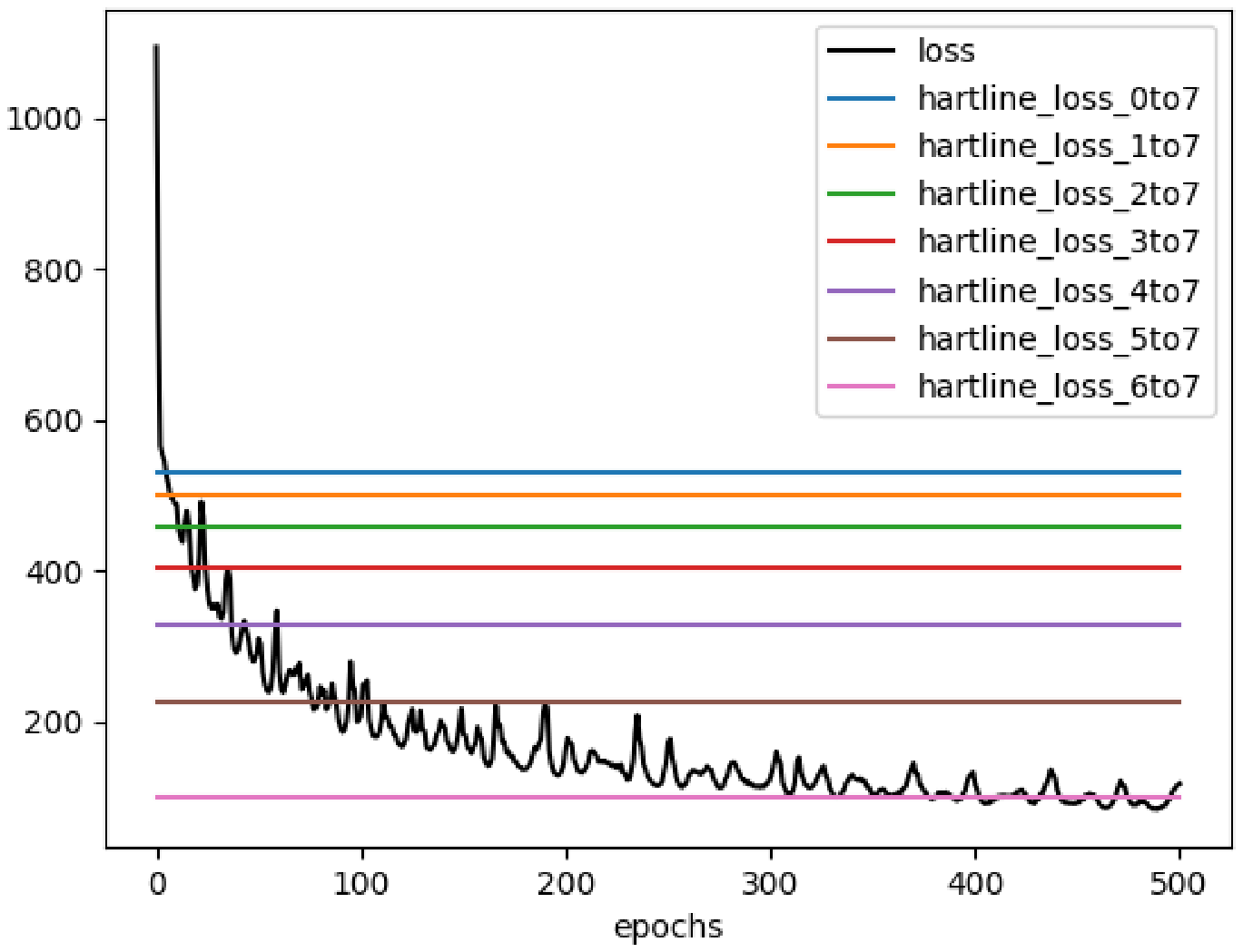}
\caption{(Left): Loss as a function of epoch $<$ 100,000; (Right): Horizontal lines show the $n$-th Hartline loss $L(X_H^{(n)}, Y) + 16.65 $ for epochs $<$ 500. }
\label{loss100,000}
\end{figure}}
The left figure in (Figure \ref{loss100,000}) shows that the loss function decreases rapidly in the first 100 epochs, and it takes finally 16.65 at epoch 100,000 as the averaged value over the last 1000 epochs. 

It is important to compare the ``Hartline's mechanism'' of the surround inhibition in physiology \cite{surround inhibition}, \cite{surround inhibition books}, with the implication obtained by DL of the neural network.  We choose six intermediate outputs, $\{X^{(1)}, X^{(2)}, \cdots, X^{(6)} \}$ during the training of DL.  The input datum is $X^{(0)}$ and the output datum is $X_{\mathrm{out}}=X^{(7)}$.  As for the Hartline mechanism, it gives the six intermediate outputs by $\{X_H^{(1)}, X_H^{(2)}, \cdots, X_H^{(6)} \}$, where $X_H^{(n)}= (\hat{H})^n X^{(0)}$, and $X_H^{(7)}= Y$ is the label.  Therefore, it is reasonable to select six intermediate epochs of the DL training so that $L(X^{(n)}, Y)=L(X_H^{(n)}, Y) + L(X_{\mathrm{out}}, Y)$ holds for $n=1, 2, \cdots, 6$, where the last term in the r.h.s fills a gap 16.65 existing between the loss functions of Hartline and DL, even at 100,000 epochs' running.  From the the training data, we select the 6 intermediate epochs.  See the right figure in (Figure \ref{loss100,000}).

The loss functions between $X^{(n)}$ and $X_H^{(n)}$, $(n=1,2, \cdots, 6)$ is a measure to understand the difference between two mechanisms,  Hartline's physiological one and DL's one.  The data shows intermediate epochs of DL which corresponds to the Hartline's $X_H^{(n)}$ are 7, 11, 18, 37, 332, for $n=1, 2, 3, 4, 5, 6$, respectively.  

The performance can be seen visually from the following two samples among 100  test data at 100,000 epoch, where the input, the label and the output signals are depicted by black, red and blue colors, respectively.  See (Figure \ref{samples}).
{\begin{figure}[h]
\centering
\includegraphics[width=80mm]{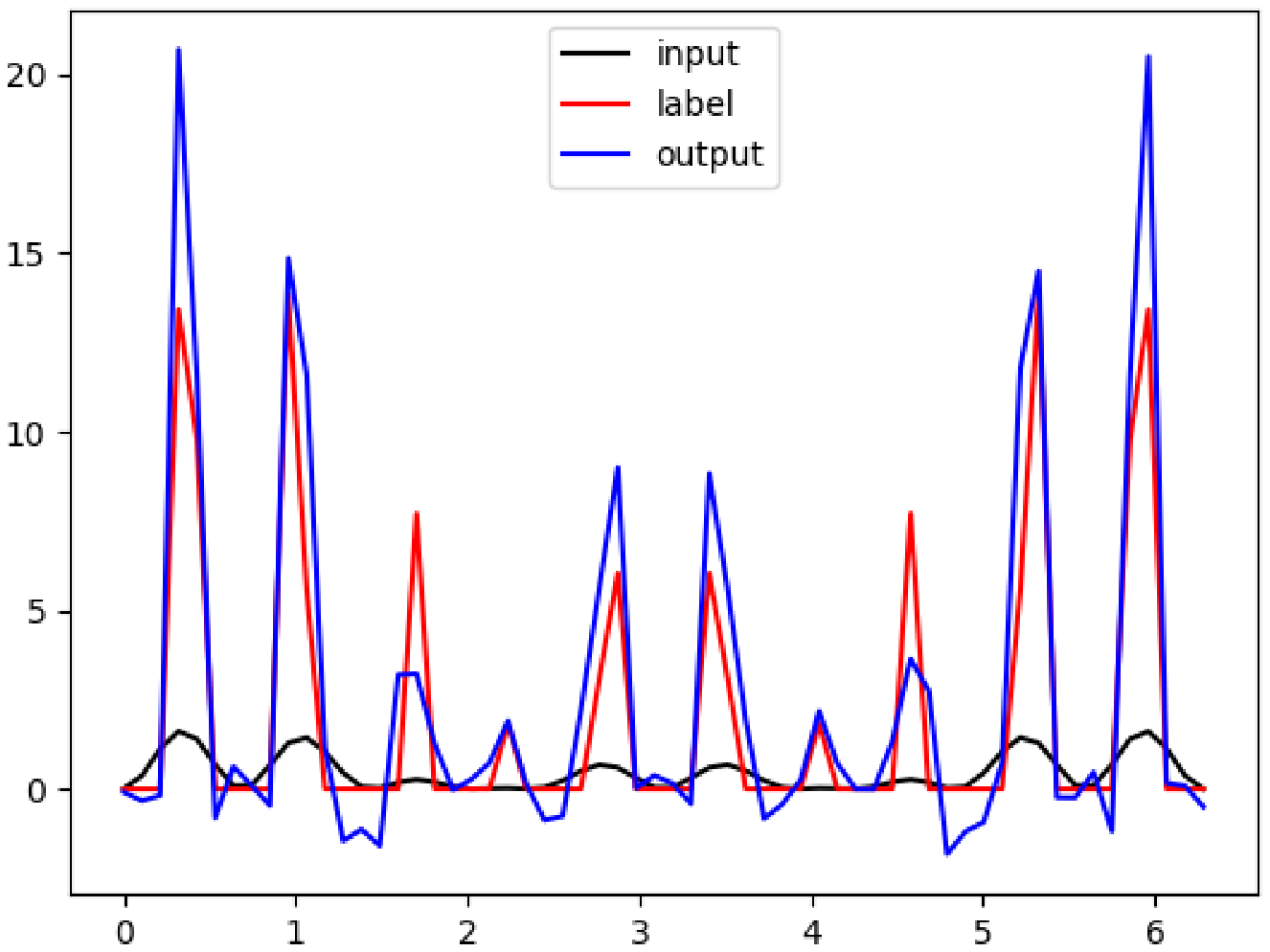}
\includegraphics[width=80mm]{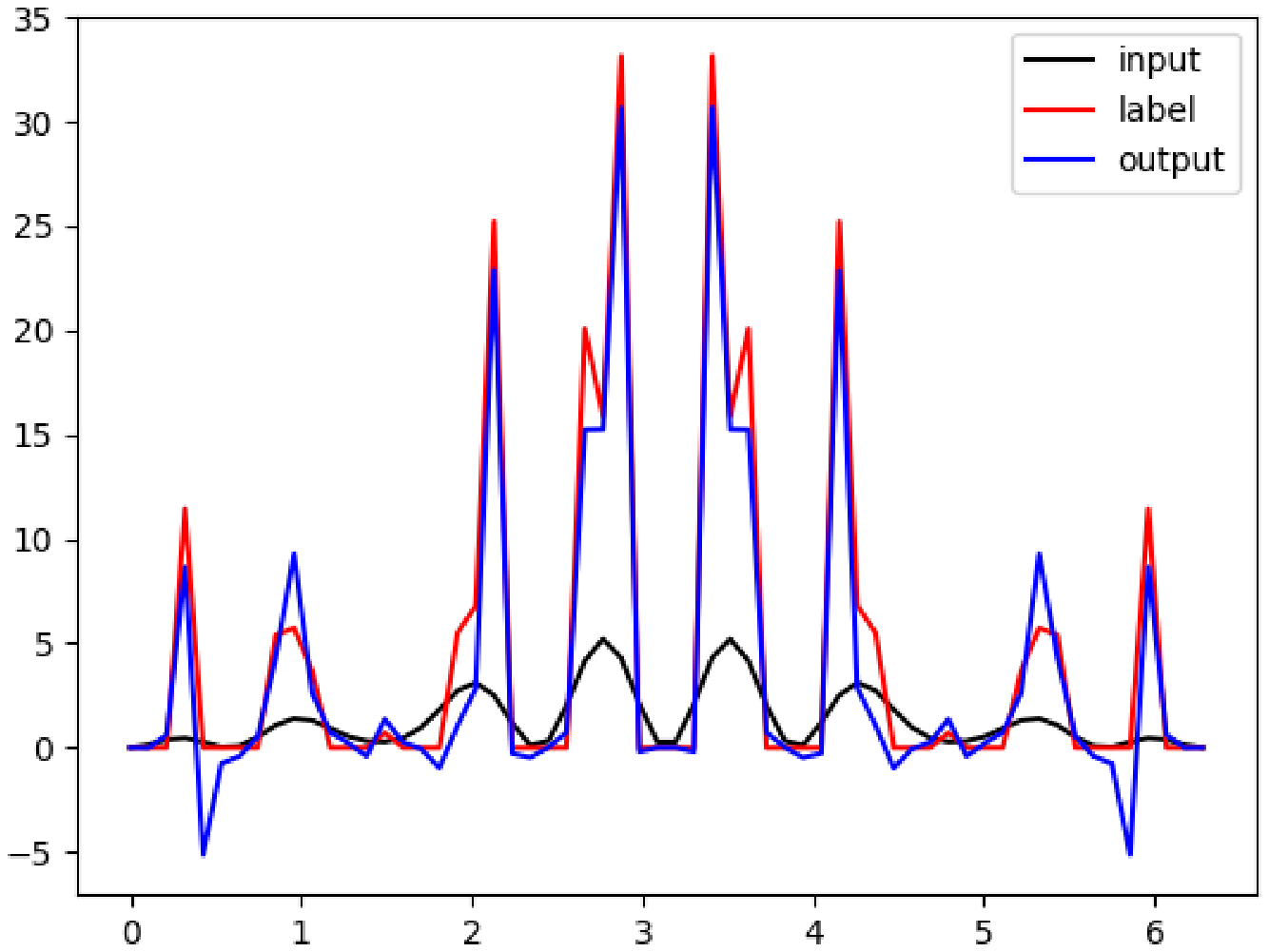}
\caption{Input (black), label (red) and output (blue) signals of two sample data among 100  test data at 100,000 epoch.} 
\label{samples}
\end{figure}}
From the samples, we can see DL acquires an ability of making the sharp contrast to the input datum, resulting the output datum in which the bumps are enhanced and strengthened.  

\section{Conclusion}

The surround inhibition mechanism of sensory nerve system is studied by deep learning of a neural network.  The deep learning mechanism by this neural network is not necessary equal to the standard one in physiology.  More detailed comparison between DL and the real sensory system is necessary, using the real data of the creature.

\section*{Acknowledgements}
We are grateful to the members of the OUJ Tokyo Bunkyo Field Theory Collaboration for fruitful discussions.



\end{document}